\documentclass[aps,twocolumn,showpacs,
floats,superscriptaddress]{revtex4}
\usepackage{graphicx}
\usepackage{dcolumn}
\usepackage{bm}
\usepackage{color}
\usepackage{amssymb}

\begin{document}
\author{B. Capogrosso-Sansone}
\affiliation{Institute for Theoretical
Atomic, Molecular and Optical Physics,
Harvard-Smithsonian Center of
Astrophysics, Cambridge, MA, 02138}

\author{\c{S}.G. S\"{o}yler}
\affiliation{Department of Physics,
University of Massachusetts, Amherst, MA
01003, USA}
\affiliation{The Abdus Salam International Centre for Theoretical Physics,
Strada Costiera 11, I-34151 Trieste, Italy}

\author{N.V. Prokof'ev}
\affiliation{Department of Physics,
University of Massachusetts, Amherst, MA
01003, USA}
\affiliation{Russian Research Center
``Kurchatov Institute'', 123182 Moscow,
Russia}

\author{B.V. Svistunov}
\affiliation{Department of Physics,
University of Massachusetts, Amherst, MA
01003, USA}
\affiliation{Russian Research Center
``Kurchatov Institute'', 123182 Moscow,
Russia}

\title{Critical entropies for magnetic ordering in
bosonic mixtures on a lattice}

\begin{abstract}
We perform a numeric study (worm algorithm Monte Carlo simulations) of ultracold
two-component bosons in two- and three-dimensional  optical
lattices. At strong enough interactions and low enough temperatures the system
features magnetic ordering.
We compute critical temperatures and entropies for the disappearance of the
Ising antiferromagnetic and the \textit{xy}-ferromagnetic order and find that
the largest possible entropies per particle are  $\sim 0.5 k_B$.
We also estimate (optimistically) the experimental hold times
 required to reach equilibrium magnetic states to be on a scale of
seconds. Low critical entropies and long hold times render the experimental
observations of magnetic phases challenging and call for increased control
over heating sources.
\end{abstract}

\pacs{67.85.Hj, 67.85.Fg, 67.85.-d}

\maketitle


\section{Introduction}
At the moment, one of the prominent focuses and major challanges of experiments
with
ultracold gases is the realization of configurations which can be used to study
quantum magnetism \cite{Sachdev_review, Lewenstein_review}. Though
interesting and fundamental on its own, better understanding of
(frustrated) magnetic systems is further motivated by its relevance to
high-$T_c$ superconductivity and applications to quantum
information processing. Direct studies of condensed matter spin systems
experimentally are limited by the lack of control over interactions, geometry,
frustration, and contaminating effects of other degrees of freedom.
A new approach consists of using ultracold atoms in optical lattices (OL)
provided that the system is driven towards regimes where it is possible
to map the corresponing (Bose-)Hubbard Hamiltonian to spin models.
\\ \indent
Striking advances in experimental techniques, e.g. high controllability and
tunability of Hamiltonian parameters, and, more recently, single site and single
particle imaging \cite{QGM_1,QGM_2,QGM_3,QGM_4}, brought forward the idea,
originally
proposed by Feynmann, of quantum simulation/emulation \cite{Feynmann}.
In the last decade, a considerable amount of theoretical and experimental
research
has been devoted to the objective of using ultracold lattice bosons and fermions
to address many outstanding condensed matter problems via Hamiltonian modeling.
Perhaps the biggest remaining experimental challenge consists of reaching low
enough temperatures/entropies for the observation of ordered magnetic states.
Theoretical insight on optimal conditions for such observations is greatly
needed. While Mott insulator (MI) phases of single component bosonic systems
have been
observed experimentally \cite{Greiner, Porto_2D, Bloch_review}, and finite
temperature effects have been extensively investigated recently
\cite{Bloch-Umass,
Pollet-Van_houcke, Ho, Capogrosso,Ketterle}, the multi-component case is
still a work in progress.
\\ \indent
In the present work, we address the issue for the case of two-component
bosonic systems. We obtain such important numbers
as critical temperatures and, more importantly entropies, below which
magnetic phases can be observed experimentally. With these numbers in hand,
we provide rough estimates of hold times required for observing thermally
equilibrated ordered magnetic states.
\\ \indent
We consider a homogeneous system of two-component bosons in a
cubic (square) lattice with repulsive inter-species interaction and half-integer
filling of each component. This system can be realized by loading OL with two
different
atomic species, see, e.g., experiments at LENS with rubidium and potassium
mixtures \cite{mixture_hetero_1,mixture_hetero_2},
or the same atomic species in two different internal energy states, see e.g
recent experiments done at MIT \cite{Ketterle} and ongoing experiments at Stony
Brook \cite{Stony_Brook}. The inter- and intra-species interaction
strengths, $U_{ab }\equiv U$,  $U_{aa}$, and $U_{bb}$ can be tuned via Feshbach
resonance
or by changing the Wannier
functions overlap (in the presence of state-dependent lattices). If the
intra-species interactions  $U_{aa}$ and $U_{bb}$ are made much larger than any
other
energy scale, and the temperature is low enough, the system is accurately
described by the
two-component \textit{hard-core} Bose-Hubbard Hamiltonian:
\begin{eqnarray}
 H=-t_a\! \sum_{<ij>}  a^{\dag}_i\,a^{}_j \, 
-t_b\! \sum_{<ij>}  b^{\dag}_i\,b^{}_j  
 \, +U\sum_{i}n^{(a)}_i n^{(b)}_i
\, . \label{hamiltonian}
\end{eqnarray}
Here $a^{\dag}_i(a^{ }_i)$,  $b^{\dag}_i(b^{}_i)$ are bosonic
creation (annihilation) operators and $t_a$, $t_b$ are hopping
matrix elements for two species of bosons ($A$ and $B$), respectively; the
symbol  $<\!\ldots\!>$ imposes the nearest-neibor 
constraint on the summation over site subscripts; $n^{(a)}_i=a^{\dag}_i a^{ }_i$
and $n^{(b)}_i=b^{\dag}_i b^{ }_i$.
\\ \indent
Model (\ref{hamiltonian}) displays a very rich ground state
phase diagram \cite{Kuklov_Svistunov,Demler_Lukin, Soyler}, see Fig.~\ref{fig1}.
For strong enough interactions, the system is incompressible in the
particle-number sector, i.e.\ it is a MI. The remaining degree of freedom
describing the boson type on
a given site can be mapped onto the effective iso-spin variable
\cite{Boninsegni,Kuklov_Svistunov,Demler_Lukin} and gives rise to two possible
MI states: a
double checker-board (2CB) solid phase, equivalent to the Ising antiferromagnet,
and a super-counter-fluid (SCF), equivalent
to a planar ferromagnet in the iso-spin terminology. For large enough
hoppings the MI state undergoes a transition to a double superfluid state (2SF).
Finally, as it has been shown recently \cite{Soyler}, for strong asymmetry
between the
hopping amplitudes and relatively weak inter-species interaction a
solid phase in the (heavy) component is stabilized via a mechanism of
\textit{inter-site} effective interactions mediated by the (light)
superfluid component.
In what follows we will focus on the magnetic states, namely the Ising
antiferromagnet and the \textit{xy}-ferromagnet.
We present the first precise results, based on path
integral Monte Carlo (PIMC) simulations by the Worm
algorithm \cite{WA}, for transition lines to magnetic phases in two- and
three-dimensions (2D and 3D) at zero and finite temperature, and
discuss experimental parameters required for reaching them.

\section{Ground state}

\begin{figure}[t!]
\includegraphics[
width=0.95\columnwidth]{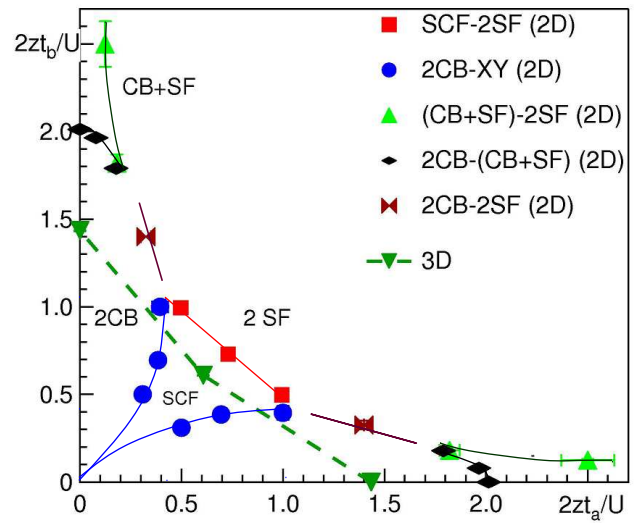}
\caption{(Colors online). Phase diagram of
model~(\ref{hamiltonian}) on a square lattice and half integer
filling factor of each component (\textit{z} is the coordination number). The
2CB-SCF first-order transition is represented by circles, the SCF-2SF second
order transition by squares, the 2CB-2SF first-order transition by
stars, the 2CB-(CB+SF) second-order transition by diamonds, and
the (CB+SF)-2SF first-order transition by up triangles. Down triangles are
benchmark points for the disappearance of magnetic order in the cubic lattice.
Lines are to guide an eye.}\label{fig1}
\end{figure}

\indent We begin with results for the ground state.  In
Fig.~\ref{fig1} we show the complete zero temperature phase diagram of model
(\ref{hamiltonian}) for the 2D system calculated in Ref.~\cite{Soyler}.
We also sketch (dashed line) the transition
line for the disappearence of magnetic order for the 3D system by computing
benchmark transition points (down triangles) for the
strongly anisotropic and isotropic limits.
These points correspond to the disappearance of the insulating Ising
and the (\textit{xy})-ferromagnetic phases, respectively. While, as expected,
the 3D case is better captured by the
mean-field theory \cite{Demler_Lukin, Soyler},
the discrepancy between mean-field and Monte Carlo results
is still sizable:  $\sim$50\%.
\\ \indent
These results provide quantitative  guidance for experimentally
achieving the regime of quantum magnetism.
In experiments with two different species this can be
done by using Feshbach resonances \cite{mixture_hetero_1} in order to reach the
desired $t_{a,b}/U$ value; in the case of the same species but
different internal states one can load state dependent
lattices and tune the interspecies interaction by changing the overlap
of Wannier functions of the two components. 

\section{Finite-temperature results}

Turning to the issue  
of reaching magnetic phases in realistic experimental setups---with an adiabatic
protocol of turning on the optical lattice---we look for highest possible
values of the critical entropy for the appearance of magnetically ordered
states.  The critical values of temperature come as a natural `by-product' of
simulations. 
In what follows we use $t_b\ge t_a$ 
as the energy unit.

\subsection{Critical temperatures}

\begin{figure}[t!]
\hspace*{0cm}\vspace*{0cm}
\centerline{\includegraphics[angle=0,width=3.0in]
{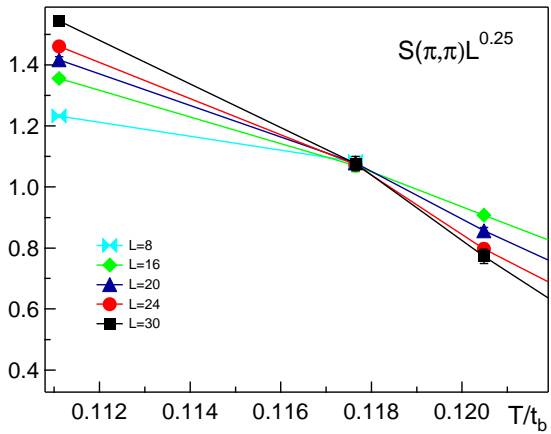}} \caption{(Colors online). Finite size scaling for the
structure factor in the 2D system (see text) for $t_a/t_b=0.285$, $U/t_b=5.7$
and system sizes $L=8, 16, 20, 24, 30$. The critical temperature can be read
form the
intersection of curves corresponding to different $L$'s. Lines are a guide to an
eye.}\label{fig2}
\end{figure}

\begin{figure}[t!]
\hspace*{0cm}\vspace*{0cm}\centerline{\includegraphics[angle=0,width=3.5in
]{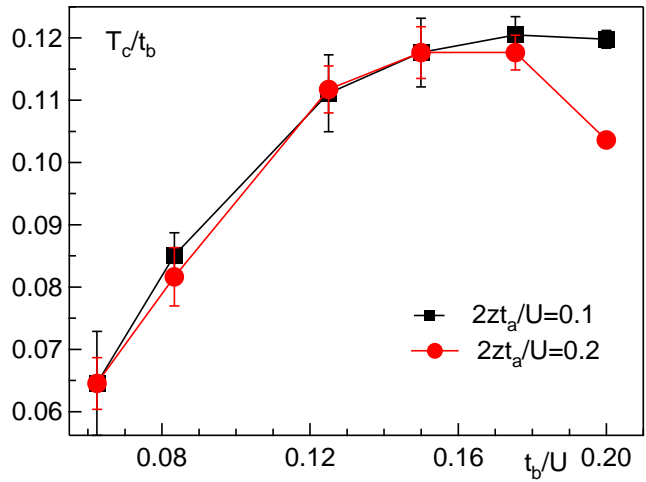}} \caption{(Colors online). Critical temperature for the Ising
state vs. $t_a/U$ in the 2D system at fixed $2zt_b/U=0.1$ and $0.2$, squares and
circles respectively.
Lines are a guide to an eye.}\label{fig3}
\end{figure}

We start with the Ising
antiferromagnet-to-normal transition. It belongs to the
\textit{d}-dimensional Ising universality class, the order parameter being the
staggered magnetization along the \textit{z}-axis or, equivalently, in bosonic
language, the structure factor (which is the square of the order parameter):
\begin{equation}
S^{(a,b)}_\textbf{\scriptsize K}  =\, \sum_{\textbf{r},\textbf{r}'}\, 
\exp\left[i\textbf{K}\!\cdot\! \left(\textbf{r}-\textbf{r}'\right)\right] \,
{ \langle n^{(a)}_{\textbf{r}} n^{(b)}_{\textbf{r}'}\rangle \over N^{(a)}N^{(b)}
}\;  ,
\label{StrF_def}
\end{equation}
with \textbf{K} the reciprocal lattice
vector of the CB solid, i.e. \textbf{K}=$(\pi,\pi)$ in 2D and
\textbf{K}=$(\pi,\pi,\pi)$ in 3D, $n^{(a,b)}_{\textbf{r}}$ the filling factor at
the site \textbf{r}, and $N^{(a,b)}$ the total number of particles A, B. In the
vicinity of the transition point, the structure factor scales as 
\begin{equation}
S_\textbf{\scriptsize K}  (\tau,
L)=\xi^{-\frac{2\beta}{\nu}}f(\xi/L)=L^{-\frac{2\beta}{\nu}}g(\tau
L^{\frac{1}{\nu}}) \; ,
\label{StrF_scaling}
\end{equation}
where $\xi$ is the correlation length, $\tau = (T-T_c)/t_b$ is the reduced
temperature, $L$
is the system size, assumed to be large enough to neglect higher-order
corrections to the universal scaling, 
$f(x)$ and $g(x)$ are universal scaling functions, and $\beta$
and $\nu$ are the critical exponent for the order parameter and correlation
length, respectively. For the 2D case $2\beta/\nu=1/4$, and for the 3D case
$2\beta/\nu=1.0366(8)$
\cite{Martin}.
At the critical point, the quantity $S_\textbf{\scriptsize K} L^{2\beta / \nu
}$ is
size independent, provided  $L$ is appropriately large, and curves of different
$L$'s intersect.
Figure~\ref{fig2} shows an example of the intersection
for the case of a 2D system, with parameters $t_a/t_b=0.285$ and $U/t_b=5.7$,
and system sizes $L=8, 16, 20, 24, 30$. The critical temperature is $T_{\rm
c}/t_b=0.1175(10)$.
\\ \indent Our results for critical temperatures in 2D are summarized in
Fig.~\ref{fig3}.
We have performed simulations at fixed $2zt_a/U=0.1, 0.2$  and varying $t_b/U$.
Our data show that the region with higher transition temperatures corresponds to
relatively
weak interactions, but away from the transition to the (CB+SF) ground
state. For strong interactions, the relevant energies, i.e. coupling of
spin degrees of freedom in the mapping to the quantum spin Hamiltonian, scale as
$\propto U^{-1}$, and
therefore require smaller temperatures in order to stabilize magnetically
ordered phases. On the other hand, for weak enough interactions,
the magnetic order will eventually disappear in favor of the (CB+SF) phase. As
we approach this transition the magnetic order becomes
weaker, therefore lower temperatures are required to observe it though the
effect 
is rather moderate. The largest transition temperatures
lie somewhere in between this two limits, and with precise numerical
simulations it is possible to accurately pinpoint
the parameter region which is best suited for current experiments.
The largest critical temperatures we have observed are  $T_c/t_b
\sim0.12$.
\\ \indent In the 3D case, we have calculated
$T_c$ in the region where we expect it to be large, $U/t_b=11$, $t_{a}/t_b=0.1$.
We have found  $T_c/t_b=0.175(15)$. The 3D simulations are far more
demanding computationally than in 2D, and the calculation of the full zero- 
and finite-temperature  phase diagram in 3D is beyond the scope of this work.
\begin{figure}[t!]
\hspace*{0cm}\vspace*{0cm}\centerline{\includegraphics[angle=0,width=3.5in
]{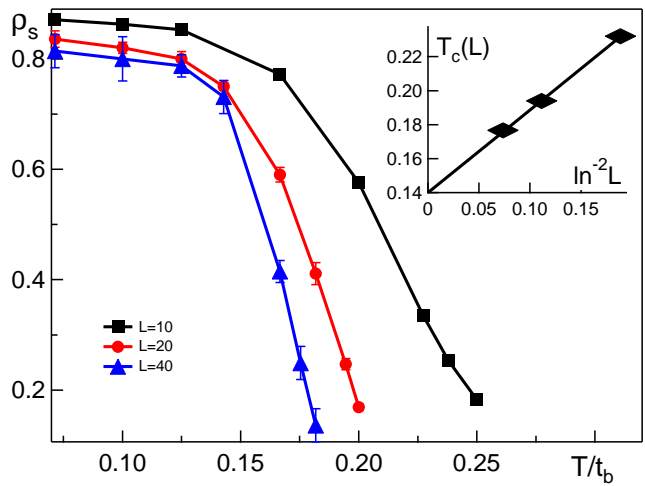}}  \caption{(Colors online). Main plot: superfluid stiffness of
the particle-hole composite object in the SCF (\textit{xy}-ferromagnetic) state
with $t_a=t_b$ and $U/t_b=11$ for system sizes L=10,20,40 trangles, circles,
squares respectively. Inset: scaling of the finite-size `critical temperature'
(see text).}\label{fig4}
\end{figure}
\\ \indent We now turn to the melting of the \textit{xy}-ferromagnetic
state. In bosonic language, it corresponds to the SCF-to-normal transition where
SCF is characterized as the superfluid state with the composite order parameter 
describing the condensate of  pairs consisting of particles of one component and
holes of the
other one,  with zero net particle flux. The transition is of the
$d$-dimensional U(1) universality class, meaning that in 2D it is of the
Kosterlitz-Thouless
(KT) type. In Fig.~\ref{fig4} we show an example of how
transition points for the 2D system are calculated. In order to
locate the critical temperature we employ finite-size arguments following from
KT renormalization-group flow for the superfluid stiffness $\rho_s$, the latter
being 
measured from statistics of fluctuations of winding numbers \cite{Ceperley}:
\begin{equation}
\rho_s =  {\langle \textbf{W}^2 \rangle \over \beta L^{d-2} }\, ,
\label{Pollock_Ceperley}
\end{equation}
where $\textbf{W}$ is the vector of worldline winding numbers in the SCF sector.
For our purposes,
it is sufficient to define $\rho_s$ up to a global pre-factor; that is why  our
Eq.~(\ref{Pollock_Ceperley})
contains no other factors.

In terms of worldline windings, the universal Nelson-Kosterlitz jump translates
into the abrupt change of $\langle\textbf{W}^2\rangle$ at the critical point
from $4/ \pi$ in the SCF phase to zero
in the normal phase. In a finite system, the universal jump is smoothed out and
winding numbers go to
zero continuously (see the main plot in Fig.~\ref{fig4}). If one defines the 
finite-size critical point $T_c(L)$ by the condition 
$\langle \, \textbf{W}^2(T_c(L))\, \rangle=4/\pi$, 
then the flow of $T_c(L)$ to the thermodynamic limit answer $T_c=T_c(\infty)$ is
given by
$T_c(L)-T_c \propto 1/(\ln L)^2$, see the inset in Fig.~\ref{fig4}.
\\ \indent We have found the following critical temperatures:
$T_c/t_b = 0.141(5)$ for $U/t_b=11$, $t_{a}/t_b=1$;
$T_c/t_b = 0.104(5)$ for $U/t_b=13$, $t_{a}/t_b=1$;
$T_c/t_b = 0.101(5) $ for $U/t_b=11$, $t_a/t_b=0.8$;
$T_c/t_b = 0.14(1)$ for $U/t_b=9.4$, $t_a/t_b=0.6$.
Critical temperatures seem to decrease as we go towards the Heisenberg point and
the effective iso-spin couplings decrease (see argument above).
Unlike the Ising-normal transition, the highest transition temperature we have
found lies close to the SCF-2SF $T=0$ transition
line. In fact, across this transition line the superfluid stiffness of the
particle-hole composites and the transition temperature to the normal state 
remain finite. As discussed in Ref.~\cite{KPS04}, at finite temperature the
SCF-2SF boundary  
moves in the direction of the 2SF ground state thus implying the following  
sequence of events: as temperature is increased 
in the vicinity of the quantum critical point the 2SF state first undergoes a
transition
to the SCF state which then turns normal at a much higher temperature. 
\\ \indent
In the 3D case, the transition point can be obtained from the finite-size
scaling of $\rho_s$. Similarly to Eq.~(\ref{StrF_scaling}), one has:
\begin{equation}
\rho_s(\tau,L)=\xi^{-1}f(\xi/L)=L^{-1}g(\tau L^{\frac{1}{\nu}}) \, .
\end{equation}
The critical temperature is extracted from the intersection of $\rho_s(\tau,L)L$
curves. We have done simulations for the system parameters $U/t_b =21$,
$t_{a}/t_b=1$ and found $T_c/t_b=0.208(7)$.
\begin{figure}[t!]
\hspace*{0cm}\vspace*{0cm} \centerline{\includegraphics[angle=0,width=3.5in
]{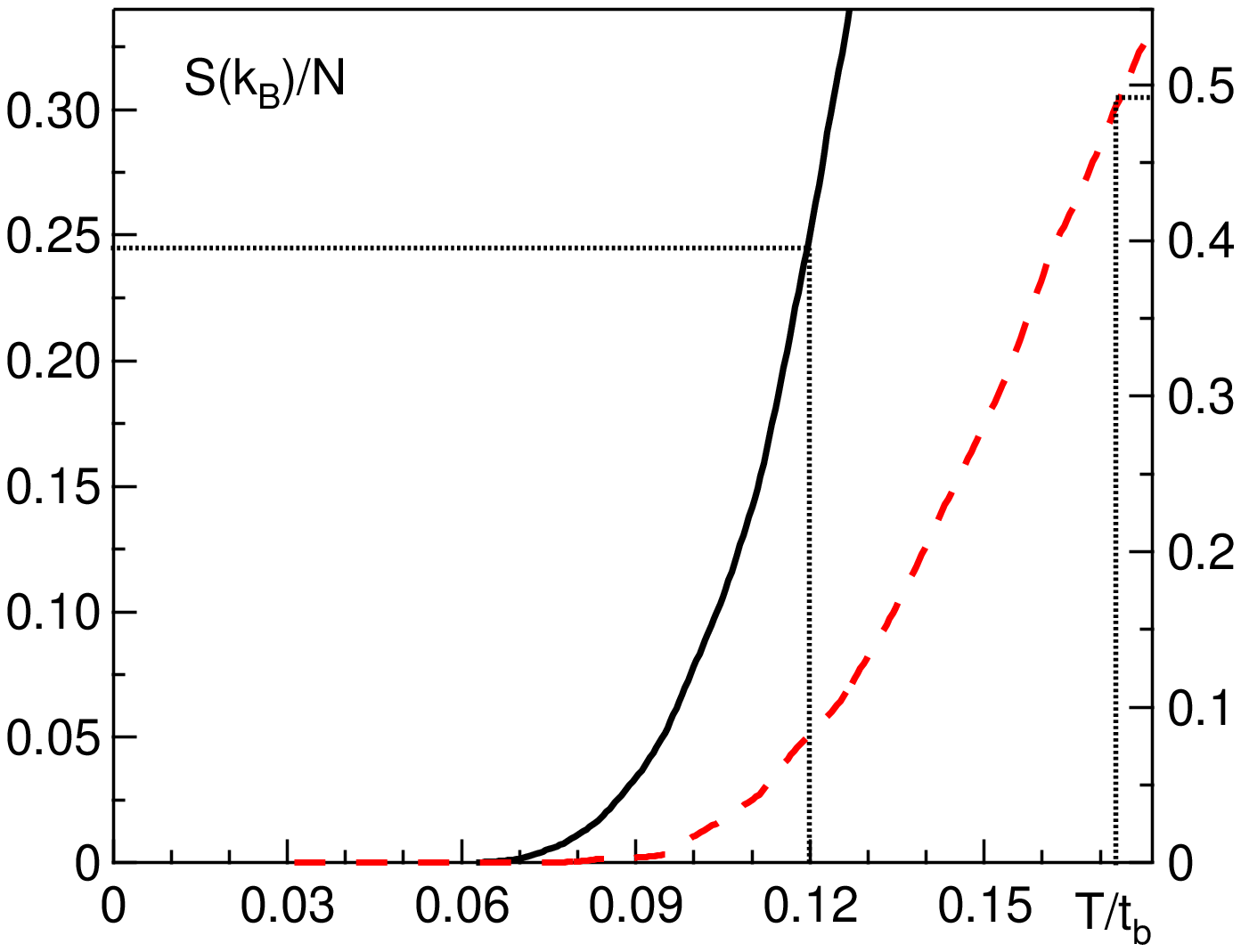}} \caption{(Colors online). Entropy curves for the Ising
antiferromagnet in 2D for $U/t_b=5.7$, $t_{a}/t_b=0.142$ and 3D for $U/t_b=11$,
$t_{a}/t_b=0.1$, 
solid and dashed lines respectively. Dotted lines
are a guide to the reading of critical entropies.
}\label{fig5}
\end{figure}
\begin{figure}[t!]
\hspace*{0cm}\vspace*{0cm} \centerline{\includegraphics[angle=0,width=3.5in
]{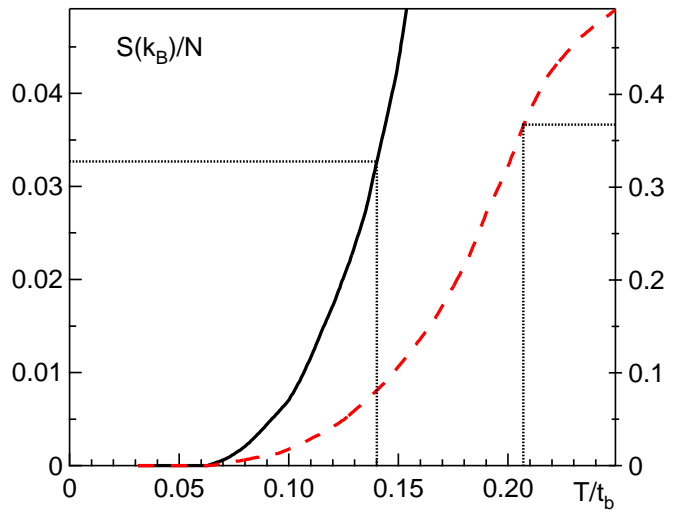}} \caption{(Colors online). Entropy curves for the
\textit{xy}-ferromagnet in 2D for $U/t_b=11$, $t_{a}/t_b=1$ and 3D for
$U/t_b=21$, $t_{a}/t_b=1$, 
solid and dashed lines respectively. Dotted lines
are a guide to the reading of critical entropies.}\label{fig6}
\end{figure}

\subsection{Entropy curves}

Entropy curves $S(T)$ are calculated starting from the energy data. We first
use spline interpolation of data points
to obtain a smooth curve $E(T)$. We then calculate entropy by using two
different numerical procedures:  (i) We obtain the specific heat $c_{\rm{V}}$ by
differentiating the spline and then calculate the entropy
by numerical integration of $c_{\rm{V}}/T$. (ii) We avoid numerical derivatives
by using  
\begin{equation}
S(T) \, =\,  {E(T)-E(0)\over T}\, +\, \int_0^T {E(T)-E(0)\over T^2}\; dT 
\end{equation}
and numerical  integration. The agreement of the two methods is very good
(within 0.5\%).
Uncertainties in entropies come therefore from the ones in critical temperatures
and finite-size effects. 
Examples of entropy curves in the
Ising antiferromagnetic state are shown in Fig.~\ref{fig5}, for
$U/t_b=5.7$, $t_{a}/t_b=0.1425$ in 2D, and $U/t_b=11$, $t_{a}/t_b=0.1$ in
3D. We find critical
entropies per particle $S_c(k_B)/N \sim 0.25\pm5\%$ and $0.5\pm20\%$ in 2D and
3D, respectively. These entropies are relatively large and definitely within the
realm
of what can be achieved with bosonic BECs.  
In Fig.~\ref{fig6} we show entropy curves for the \textit{xy}-ferromagnetic
state. 
The critical entropy
in 2D for $U/t_b=11$, $t_{a}/t_b=1$ is $S_c(k_B)/N\sim 0.033\pm5\%$, about an
order of magnitude
smaller (!!) than for the 3D value $S_c(k_B)/N\sim0.35\pm10\%$ obtained for 
$U/t_b=21$, $t_{a}/t_b=1$.  
This is explained by the specifics of the KT transition when the SF density
jumps to zero
discontinuously at the critical point, i.e. when the system thermodynamics is
still dominated
by the dilute phonon gas. Correspondingly, at the transition temperature the
thermal 
energies and entropies are low. Our thermodynamic data confirm that this is
precisely 
what is happening for the 2D system: energy scales with temperature as $\propto
T^3$  (which implies that entropy is $\propto T^2$) all the way up to
temperatures $T<T_c$.

\section{Minimal experimental hold times}  

Finally, we  estimate minimal hold times required to observe ordered magnetic
phases 
under typical experimental conditions.
\\ \indent For a cubic lattice and using a harmonic approximation around the
minima of the optical lattice potential \cite{Jaksch}, the tunnelling matrix
elements
and  on-site interaction energies are given by:
\begin{equation}
t_{a,b}\, \approx\, 
\frac{4}{\sqrt{\pi}} \left(E_R^{(a,b)}V_{a,b}^3 \right)^{1\over 4} {\rm exp}
\left(-2\sqrt{V_{a,b} / E_R^{(a,b)}} \right) ,
\label{hopping_eq}
\end{equation}
\begin{equation}
U\, \approx\, 
\frac{4\sqrt{\hslash}}{\sqrt{\pi}}\, a_s^{(ab)} \,
\overline{m\omega}^{3/2}\frac{1}{2\nu_
{
ab}} \; ,
\label{interspecies_eq}
\end{equation}
\begin{equation}
U_{a,b}\, \approx\, 
\frac{\sqrt{2\hslash}}{\pi}\,
a_s^{(aa,bb)}(m_{a,b}\omega_{a,b})^{3/2}\frac{1}{m_{ab}}
\; ,
\label{intraspecies_eq}
\end{equation}
where
\begin{equation}
\overline{m\omega}=\frac{m_a\omega_am_b\omega_b}{m_a\omega_a+m_b\omega_b} \, ,
\end{equation}
and 
\begin{equation}
\omega_{a,b}=\sqrt{4E_R^{(a,b)}V_0^{a,b}}/\hslash
\end{equation}
 is the harmonic oscillator frequency, 
 \begin{equation}
 E_R^{(a,b)}=\frac{\hslash^2 k^2}{2m_{a,b}}
 \end{equation}
is the atomic recoil energy, $m_{a,b}$ and $\nu_{ab}$ are the bare and
reduced masses respectively, $a_s^{(aa,bb)}$ and $a_s^{(ab)}$
are the intra- and interspecies scattering lengths.
The hard core limit can be achieved if e.g. $a_s^{(aa,bb)} < a_s^{(ab)}$, or by
manipulation of the overlapping of Wannier functions as explained above.
For $^{87}$Rb -$^{41}$K mixtures \cite{mixture_hetero_1} and away from
resonances one has
$a_{\rm Rb-K}=163a_0$, $a_{\rm Rb}=99a_0$, and $a_{\rm K}=65a_0$ ($a_0$ is the
Bohr
radius). One can then use Feshbach resonances to tune scattering lengths to the
hard-core limit.
\\ \indent
To estimate the hold time $t_{\rm exp}$ required for the
observation of the magnetic phases we look at the lowest dynamic 
energy scale in the system which is $t_a$ in our case.
Clearly, unless a condition $t_{\rm exp} \gg h / t_a$ is satisfied, one
may not even discuss thermally equilibrated normal states, not to mention
low temperature ordered ones. If $T_c$ is smaller than $t_a$, we consider 
$t_{\rm exp}\gg h/T_c$ as the minimal requirement.
As we have seen, the optimal experimental parameters for both Ising and 
\textit{xy} phases result in ${\rm min}\, (t_a,T_c)\sim 0.1 t_b$, and in what
follows we will
use this energy scale for the estimate of the hold time.     

Let us consider laser beams with $\lambda=$1064 nm and discuss
the mixtures of Rb atoms in states $\mid\! \!1,-1\, \rangle $ and $\mid\!\!
2,-2\, \rangle $ \cite{Ketterle,
Stony_Brook}, for which $a_s^{(ab)}=98.09a_0$.
For the melting of the Ising state we require $U/t_b \sim 11$ and
$t_a/t_b \sim 0.1$ which translates into the optical lattice depths
$V_a/E_R^{(a)} \sim 19.5$ and $V_b/E_R^{(b)} \sim 9$, and the final result
$t_{\rm exp}\gg 0.2$s.
For the melting of the \textit{xy}-ferromagnet we require $U/t_b \sim 21$, 
$t_a\sim t_b$, or, in terms of the lattice depths, $V_a/E_R^{(a)}=V_b/E_R^{(b)}
\sim 12$,
which
implies that $t_{\rm exp}\gg 0.035$s. 
For the case of $^{87}$Rb -$^{41}$K mixtures, the best-case scenario corresponds
to $a_{\rm Rb-K}=163a_0$ and $a_{\rm Rb},a_{\rm K}\gg a_{\rm Rb-K}$ achievd via
Feshbach
resonances for intraspecies collisions. We consider the \textit{b} species to be
$^{87}$Rb.
A similar analysis of the Ising antiferromagnetic case 
leads to $V_a/E_R^{(a)} \sim 19.5$, $V_b/E_R^{(b)}  \sim 6$ and 
$t_{\rm exp}\gg0.08$s. For the \textit{xy}-ferromagnetic case 
we have $V_a/E_R^{(a)} \sim 11.5$ and $V_b/E_R^{(b)} \sim 8.6$ and 
$t_{\rm exp}\gg0.015$s. If, instead, one tunes the interspecies scattering
length to, e.g., 
$a_{\rm Rb-K}\sim 35a_0$, in order to achieve the hard-core limit, this implies
$V_a/E_R^{(a)} \sim 26.2$, $V_b/E_R^{(b)} \sim 10.6$, 
$t_{\rm exp}\gg0.25$s for the Ising antiferromagnet, and $V_a/E_R^{(a)} \sim
17.3$, $V_b/E_R^{(b)} \sim 13.6$, 
$t_{\rm exp}\gg0.05$s for the \textit{xy}-ferromagnet.
\\ \indent From these estimates we conclude that observing ordered 
magnetic phases will be experimentally challenging since the required 
time scales might have to exceed seconds (with some advantage for dealing with
the  
$^{87}$Rb -$^{41}$K mixture). Increasing the sample stability and suppressing
various
heating mechanisms (three-body losses, background vacuum, spontaneous scattering
of lattice photons, 
and technical noises such as beam alignment, intensity fluctuations, mechanical
vibrations) 
has to be achieved. To appreciate the problem, we mention the 
 heating rate (entropy per particle) of $\sim 1k_B/$s observed recently  in a
typical experiment
 in the optical lattice \cite{Bloch-Umass}.

\section{Conclusion}  

We have  addressed numerically (by worm algorithm Monte Carlo simulations)  the
problem of  magnetic ordering in the two-component
Bose-Hubbard model in the intraspecies hard-core limit, for 2D and 3D cases, at
finite
temperature. The emphasis of the study is on revealing the optimal parameters 
for (and analyzing the feasibility of) experimentally 
achieving the transitions to Ising antiferromagnetic (a.k.a.\ checkerboard
solid) and $xy$-ferromagnetic (a.k.a.\ super-counter-fluid)
phases. We have identified the optimal experimental conditions, corresponding to
maximal critical entropy per particle.
temperatures and entropies. On the basis of our data, we have estimated minimal
experimental hold times
required to reach equilibrium magnetic states. These times have to be on a scale
of
seconds which renders the experimental observations of magnetic phases
challenging and calls for increased control
over heating sources.

Our results---optimal Hamiltonian parameters with corresponding values of
critical entropies, temperatures, and minimal hold times---can be directly used
for guiding 
and benchmarking the on-going experiment on creating optical lattice emulators.
\\
\\
\indent
We would like to thank D. Schneble, D. Pertot, B. Gadway, F. Minardi, M.
Inguscio, J. Catani, G. Lamporesi, G. Barontini, G. Thalahammer, W.
Ketterle, D. Weld, H. Miyake for fruitful discussions.
This work was supported by ITAMP, DARPA OLE program and the NSF grant
PHY-0653183.

\end{document}